# Web Portal for Photonic Technologies Using Grid Infrastructures

Hrachya Astsatryan[1], Tigran Gevorgyan[2] and Anna Shahinyan[3]

[1]Institute for Informatics and Automation Problems, National Academy of Sciences, Yerevan, Armenia; [2]Institute for Physical Researches, National Academy of Sciences, Ashtarak, Armenia; [3]Yerevan State University, Yerevan, Armenia.
Email: hrach@sci.am, t_gevorgyan@ysu.am, anna_shahinyan@ysu.am

**ABSTRACT**

The modeling of physical processes is an integral part of scientific and technical research. In this area, the Extendible C++ Application in Quantum Technologies (ECAQT) package provides the numerical simulations and modeling of complex quantum systems in the presence of decoherence with wide applications in photonics. It allows creating models of interacting complex systems and simulates their time evolution with a number of available time-evolution drivers. Physical simulations require massive amounts of calculations are often executed on distributed computing infrastructures. It is often difficult for non expert users to use such computational infrastructures or even to use advanced libraries over the infrastructures, because they often require being familiar with middleware and tools, parallel programming techniques and packages. The Parallel Grid Run-time and Application Development Environment (P-RADE) Grid Portal is a Grid portal solution that allows users to manage the whole life-cycle for executing a parallel application on the computing Grid infrastructures. The article describes the functionality and the structure of the web portal based on ECAQT package.

## 1. Introduction

Recent years the greatly evolved area of physics is modeling of physical processes, which is a result of progress in computer science and its effectiveness in the means of technologies. Theoretical results obtained within framework of computer modeling allow to have full view of systems and processes as more complicated systems do not have analytical solutions and require approximations.

In this paper, a Web portal for photonic and quantum technologies, using computational Grid infrastructures, is introduced that makes possible to solve various problems for quantum open systems. Such systems are mainly constructed from photons, including laser fields, atom-photon interactions and nonlinear optical interactions in crystals and describe photonic devices in the presence of imperfection effects: quantum noise, decoherence and etc. Up to now, the modeling of photonic devices on the level of their practical applications is not yet well standardized. On the other hand, progress in this area open up exciting possibilities for exploring fundamental physics as well as applications in telecommunications, signal processing and quantum information technology. The work is the logical continuation of previous works [1,2], with a target to provide a research platform, which will use transparency large amount of computational resources of Grid infrastructure without being aware of where input, output data and computational resources are located. This is very important, because distributed computing infrastructures, particularly Grids are complicated and therefore these are difficult for scientists and researchers to use without advanced knowledge and experience.

The application has been constructed on the base of the numerical simulation method of quantum trajectories or Quantum State Diffusion (QSD) method [3]. The corresponding library includes investigation of quantum dissipative chaos, bistability and bifurcations [4-8], quantum stochastic resonance [9], long-lived quantum interference in periodically modulated oscillatory systems [10], engineering of Fock states and qubits in nonlinear oscillators [11] elaboration of devices generating intensive entangled light beams for quantum communications [12-15] as well as production of three-photon entangled states in parametric devices [16-18] and numerical simulation of complex quantum optical systems (see, particularly, interaction of an atom with bichromatic laser field [19-21], cascaded processes [22] and photonics in ion-trap systems [23]).

The application handles systems, which satisfy bosonic quantum statistics. In the current implementation the bosonic operators cover only part of the possible mater-light interactions and elaboration of the package involving fermion operators is now in the progress. It should be mentioned the other papers in the field of numerical simulation of open quantum systems based on the library of QSD approach [24–26].

Behavior of physical objects can be explained and predicted with system of differential equations. In recent years quantum physics has taken on special significance and wide applications in which it is no longer possible to neglect the environment interaction: dissipation and decoherence. In this way, the corresponding quantum systems are usually treated as open systems and their time-evolution is described by a density matrix in the framework of the master equation but not on the base of a Hilbert-space vector $|\Psi\rangle$ and the Schrodinger equation. In general, for the most physical systems the master equation cannot be solved analytically for arbitrary evaluation times and need to use numerical methods. But even a numerical solution of the master equation can be very hard. If a state vector $|\Psi\rangle$ requires $D$ basis vectors in Hilbert space to represent it, the corresponding density operator will require $D^2 - 1$ real numbers. This basis can often be too large even for the most powerful machines to handle, particularly, in the case if the system involves more than one degree of freedom. This problem can be overcome by unraveling the density operator evolution into quantum trajectories [3]. Since quantum trajectories represent the system as a state vector rather than a density operator, they often have a numerical advantage over solving the master equation directly, even though one has to average over many quantum trajectories to recover the solution of the master equation.

One of widely used approximations for open quantum systems is the Markovian dynamics described in terms of the Lindblad master equation for the reduced density matrix $\rho$. The time-evolution in this approach is governed by the Hamiltonian operator $H$ and the Lindblad operators of the system. Thus, in the current implementation QSD method is used based on the stochastic equation that involves both Hamiltonian and the Lindblad operators for the state $|\Psi(t)\rangle$. The density operator using an expansion of the state vector $|\Psi\rangle$ in a truncated basis of Fock's number states of a harmonic oscillator (photonic states) is calculated. In this way, the ECAQT library corresponding to QSD method has been applied for various physical problems [4–18].

In this paper, our previous results by developing a user-friendly gridified platform based on Grid portals are expanded that are very popular for designing user interfaces to Grids. The Grid portal build upon the familiar Web portal model offer to virtual communities of users a single point of access to computational or data resources. The P-GRADE [27] Grid Portal is a Grid portal solution that allows users to manage the whole life-cycle for executing a parallel application on the Grid, enabling the creation, execution and monitoring of work flows through high-level Web interfaces. Being built onto the GridSphere portal framework, the P-GRADE Portal hides the low level details of Grid systems with high-level, user-friendly interface that can be easily integrated with various middleware. The single sign-on feature of the portlet allows users to access Grid resources by automatically retrieving proxy credentials for users. The experiments have been carried out using computational resources of the Armenian National Grid infrastructure [28] that consists of seven Grid sites located in the leading research and educational organizations of Armenia. The infrastructure is a part of the pan-European Grid infrastructure within the European Grid Initiative. Mainly the "ARMGRID.GRID.AM" multidisciplinary virtual organization (VO) [29] is used, because the applications do not require any special software components to run over the infrastructure. This paper is organized as follows. Section 2 describes the structure of the suggested Web Portal. Section 3 introduces the parallelization algorithm and benchmarking results. Finally, Section 4 presents our conclusions.

## 2. Web Portal Structure

This Grid aware portal (see Fig. 1) is the logical continuation of our previous work for developing a Grid-aware Web interface for quantum optics technologies. (see Fig.2).
Mainly the authorization, authentication, job management and information services of the PGrade Portal have been used by the portlet. The single sign-on feature of the portlet allows users to access Grid resources by automatically retrieves proxy credentials for users. The Web Portal consists of a few layers (see Fig. 3). In the first step user chooses the requested process or the combination of processes, that needs to prepare and to submit the algorithms for calculations of based quantities for Quantum Optics (the density matrix, the quasi-distributions in that number the Wigner function in phase-space, and the fidelity for identification of quantum states); the Poincaré section describing dissipative chaos in semi-classical, and the entropy as a measure in information theory. Each simulation algorithm independent from the others and also runs separately. Thus, it makes possible to sign all these four algorithms as well as one of them. The mean problem for open quantum system is the calculation of the density matrix which leads to calculations of the other quantities describing the system. For choosing the Density Matrix algorithm, on the next step user needs to concretely denote the system by giving the input parameters: the system's Hamiltonian (creation and annihilation operators, environment operators, values of the external laser field, nonlinearity of media and etc.), system's freedom numbers, system's time evolution range, output time intervals, number of trajectories, number of nodes for parallel calculation and etc. Note that in this case the other remained algorithms are getting automatically. In the case of choosing the other remained algorithms without choosing Density Matrix algorithm, on the next step, user needs additionally to introduce the input files-density matrices. In the last step (job submission space) user should choose the output files formats, location of these files and the time interval for getting output runtime if a run time is available for this calculation. On the last step all this parameters are given with default values. The job submission module is the main module interacts between user and infrastructure. The module consists of the following services:

- Myproxy, which allows a grid user to delegate her/his proxy certificate;
- Job matchmaker, that finds a suitable resource within the VO that matches the job requirements;
- Converter that converts all given information into the language for submission, (job description language service (JDL));
- Job management service that stores all information into the mysql database, submit the job and check the status.

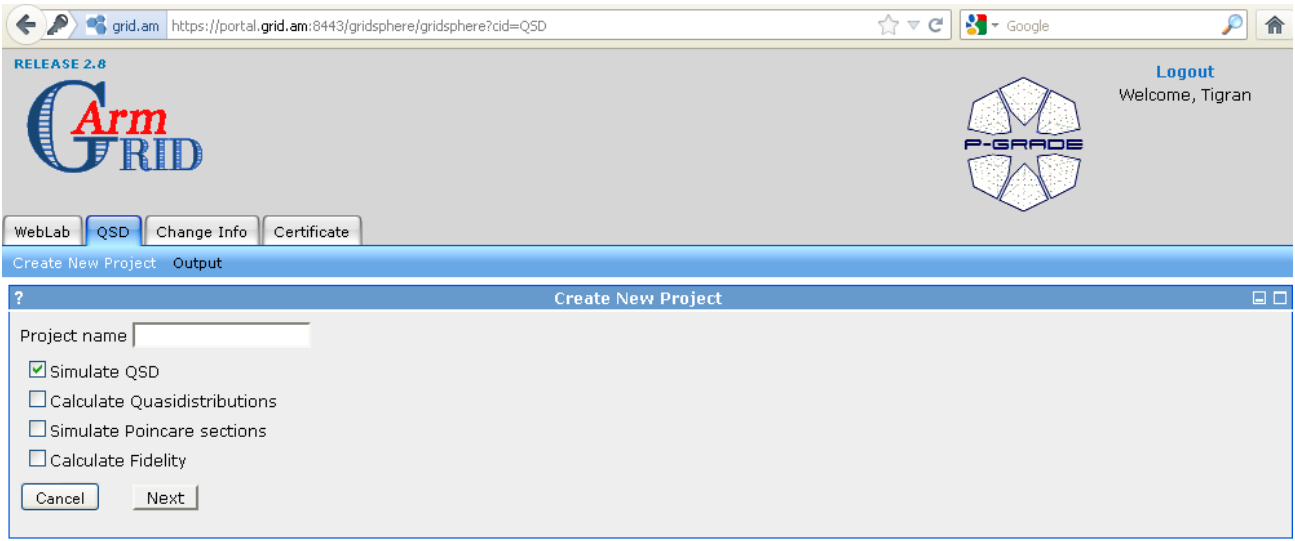

**Fig. 1** Screenshot of the P-GRADE Web Portal Interface.

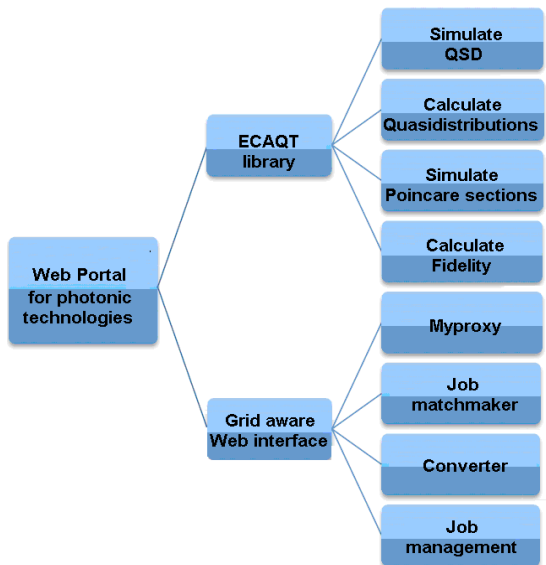

**Fig. 2** The structure of the Web portal for photonic technologies.

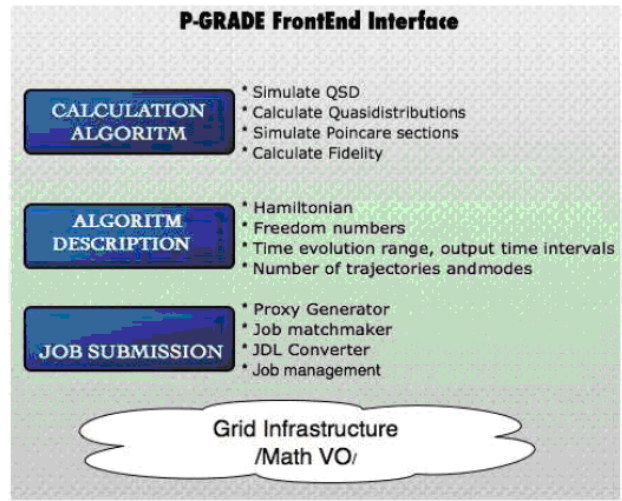

**Fig. 3** Web Portal layers interconnections within the environment.

Balancing the calculations over a heterogeneous computing infrastructure such as the ArmGrid is quite challenging. The job matchmaker service makes a decision based experimental results [30] using all platforms available in the ArmGrid infrastructure. The converter service makes the JDL file by indicating quantity of CPUs and size of RAMs as requirements.

After submitting the job with the help of the specific mediator object which is written in Java, job submitted requested Grid site indicated previously. This mediator object is used to establish the connection between Web Portal and calculation algorithms.

The calculation algorithms are based on ECAQT library [1-2], which is given in the first step on Web Portal. The ECAQT library software package is relied on generic programming and template meta-programming. In particular, the principal concept of the design of library is that all information which gets from Web Portal available at compile time should be processed as such, with the help of Mediator layer. Its design keeps abstraction and multiple code usage. ECAQT consists of three layers, GUI written in Qt, Mediator layer, Engines. The Mediator layer is created to make engines indepenendent from GUI. This allows switching GUI implantations without changing Engines. Mediator is the interface between GUI and Engines. The Engines layer consists of four cores: QSD, Poincare section calculation, quasidistributions, fidelity calculations. The details can be found in [1, 2]. The ECAQT is called extendible as it allows changing and adding the application behavior for C++ programmers from GUI. The user can specify own format to read the input file for some engines. Particularly Fidelity engine is fully open for extensions. It has a library for pure states that can be extended by user. To extend the pure states library user should inherit from states::state class and overload generate norm() and generate state in fock basis() virtual functions. The only restriction in the current implementation is that the state should be represented in Fock basis.

After job submission the input parameters are parsed and source code is compiled on the specified Grid site. After parsing and compilation process the existing errors are available in an error file. If there are no error messages in submission page in the Web Portal appears a message: "job is submitted". All output information is stored in a database. Information about current job status (running, aborted or finished) and also user ID, dates of submitting, input parameters value, output files are available on the monitoring page of Web Portal. This monitoring page is grouped all information history about user's jobs and makes it visible graphically.

## 3. Parallelization Algorithm

As a rule, numerical simulations for the systems which contain more than one dimensions, or the numbers of freedom, take long time. For example, in area of photonics state vectors of such systems contain about 50000 complex numbers corresponding of each QSD trajectory that takes minutes. Since at least 1000-5000 trajectories are required to get smooth results a full simulation takes days depending on the system freedom degree. The suggested QSD parallel algorithm simulates the same equation many times with the same initial state and exactly with the same parameters in order to reduce execution time. The algorithm has been realized using Massage Passing Interface (MPI) and has no limitation on specified processors number. We used Boost library MPI interface. This allows to use complex types in MPI. In order to use the complex types in Boost MPI interface, these types should be serialized according to Boost serialization library.

The algorithm is scalable and efficient by avoiding some problems like synchronization and time delay in massage passing processes in inter-processors communication.

In the first stage the main.cpp module calling the run() function in order to takes the input parameters including the number of trajectories (n = 5000) and the number of computational nodes ($N$). The MPI *Comm_rank()* function returns the number of rank, which get *run()* function as a parameter. While working in parallel mode the equal number of trajectories is being simulated on different nodes ($1$-$N$) of the Grid site. After simulation the results from different processes are merged. In this way a runtime of the simulation can be reduced on many times, that explicitly depends on number of available nodes in the indicated Grid site. It is more significant that parallelization in this approach provides linear effectiveness to solve the master equation.

The procedure of parallelization has been performed as for the density matrix as well as for the arbitrary quantities of interest including the Wigner function that is a complex two-dimensional function visualizing quantum states in a phase-space. As we are interested in time-evolution of the system our aim is to calculate the Wigner functions for different time points during a given time interval. Thus, the corresponding calculation layer in our package is also parallelized in the same way. We run the program with input files for different times using different nodes.

## 4. Conclusions

The full implementation of the suggested environment for numerical simulation of photonic technologies over the Grid will allow users, who are not familiar with the parallel programming technologies and software tools (Grid and Cluster middlewares, MPI, JAVA, Unix OS, Open PBS, Condor, etc.), to create and submit their requests over the portal that hides all details of the underlying distributed computing infrastructure. The portal uses open source technologies and software tools and can be easily integrated with any Grid infrastructure and a VO that supports corresponding libraries.

## 5. Acknowledgements

We would like to acknowledge G. Yu. Kryuchkyan for productive discussions while developing user interface. We also would like to acknowledge Boost, Open MPI developers.

## REFERENCES

[1] H. H. Adamyan, N. H. Adamyan, N. T. Gevorgyan, T. V. Gevorgyan, and G. Yu. Kryuchkyan, "Software for numerical simulations in the field of quantum technologies based on parallel programming", Physics of Particles and Nuclei Letters 5, N3, 161 (2008). doi:10.1134/S1547477108030047

[2] T. V. Gevorgyan, A. R. Shahinyan, G. Yu. Kryuchkyan, "Extendible C++ Application in Photonic Technologies based on Parallel Computing", CSIT proceedings 2009, Yerevan, 27 Sept.-2 Oct., p. 367, (2009).

[3] N. Gisin and I. C. Percival, "The quantum-state diffusion model applied to open systems", J. Phys. **A25**, 5677 (1992); I. C. Percival, Quantum State Diffusion (Cambridge University Press, Campridge, (2000). doi:10.1088/0305-4470/25/21/023

[4] S. T. Gevorkyan, G. Yu. Kryuchkyan, and N. T. Muradyan, "Quantum fluctuations in unstable dissipative systems", Phys. Rev. **A61**, 043805 (2000). doi:10.1103/PhysRevA.61.043805

[5] H. H. Adamyan, S. B. Manvelyan, and G. Yu. Kryuchkyan, "Chaos in a double driven dissipative nonlinear oscillator", Phys. Rev. **E64**, 046219 (2001). doi:10.1103/PhysRevE.64.046219

[6] G. Yu. Kryuchkyan and S. B. Manvelyan, "Quantum Dissipative Chaos in the Statistics of Excitation Numbers", Phys. Rev. Lett. 88, 094101 (2002). doi:10.1103/PhysRevLett.88.094101

[7] G. Yu. Kryuchkyan and S. B. Manvelyan, "Sub-Poissonian statistics in order-to-chaos transition", Phys. Rev. **A68**, 013823 (2003). doi:10.1103/PhysRevA.68.013823

[8] T. V. Gevorgyan, S. B. Manvelyan, A. R. Shahinyan, and G. Yu. Kryuchkyan, "Dissipative Chaos in Quantum Distributions", "Modern Optics and Photonics: Atoms and Structured Media". Eds: G. Kryuchkyan, G. Gurzadyan and A. Papoyan, World Scientific, 60-77, (2010). doi:10.1142/9789814313278_0005

[9] H. H. Adamyan, S. B. Manvelyan, and G. Yu. Kryuchkyan, "Stochastic resonance in quantum trajectories for an anharmonic oscillator", Phys. Rev. **A63**, 022102 (2001). doi:10.1103/PhysRevA.63.022102

[10] T. V. Gevorgyan, A. R. Shahinyan, G. Yu. Kryuchkyan, "Quantum interference and sub-Poissonian statistics for time-modulated driven dissipative nonlinear oscillators", Phys. Rev. **A 79**, 053828 (2009). doi:10.1103/PhysRevA.79.053828

[11] T. V. Gevorgyan, A. R. Shahinyan, G. Yu. Kryuchkyan, "Generation of Fock states and qubits in periodically pulsed nonlinear oscillators", Phys. Rev. **A 85**, 053802 (2012). doi:10.1103/PhysRevA.85.053802

[12] H. H.Adamyan, N. H.Adamyan, S. B.Manvelyan, and G.Yu.Kryuchkyan, "Quadrature entanglement and photon-number correlations accompanied by phase-locking", Phys. Rev. **A73**, 033810 (2006).

[13] H. H. Adamyan and G. Yu. Kryuchkyan, "Time-modulated type-II optical parametric oscillator: Quantum dynamics and strong Einstein-Podolsky-Rosen entanglement", Phys. Rev. **A74**, 023810 (2006). doi:10.1103/PhysRevA.74.023810

[14] N. H. Adamyan, H. H. Adamyan and G. Yu. Kryuchkyan, "Time-domain squeezing and quantum distributions in the pulsed regime", Phys. Rev. **A77**, 023820 (2008). doi:10.1103/PhysRevA.77.023820

[15] G. Yu. Kryuchkyan and L. A. Manukyan, "Entangled light in transition through the generation threshold", Phys. Rev. **A69**, 013813 (2004). doi:10.1103/PhysRevA.69.013813

[16] D. A. Antonosyan, T. V. Gevorgyan, and G. Yu. Kryuchkyan, "Three-photon states in nonlinear crystal superlattices", Phys. Rev. **A 83**, 043807 (2011). doi:10.1103/PhysRevA.83.043807

[17] G.Yu. Kryuchkyan, N.T. Muradyan, "Toward the multiphoton parametric oscillators", Phys. Lett. **A 286**, 113 (2001). doi:10.1016/S0375-9601(00)00801-X

[18] G. Yu. Kryuchkyan, L. A. Manukyan, N. T. Muradyan, "Three-photon light in repeated photon splitting". Optics Communication, 190, 245 (2001). doi:10.1016/S0030-4018(01)01095-1

[19] M. Jakob, G. Yu. Kryuchkyan, "Squeezing in the resonance fluorescence of a bichromatically driven two-level atom", Phys. Rev. **A58**, 767 (1998). doi:10.1103/PhysRevA.58.767

[20] G. Yu. Kryuchkyan, M. Jakob, A. S. Sargsian, "Resonance fluorescence in a bichromatic field as a source of nonclassical light", Phys. Rev. **A57**, 2091 (1998). doi:10.1103/PhysRevA.57.2091

[21] M. Jakob, G.Yu. Kryuchkyan, "Autler-Townes effect with mono- and bichromatic – pump fields: Reservoir effects and Floquet-state treatment", Phys. Rev. **A57**, 1355 (1998). doi:10.1103/PhysRevA.57.1355

[22] K. V. Kheruntsyan, G. Yu. Kryuchkyan, N. T. Mouradyan, K. G. Petrossian, "Controlling instability and squeezing from a cascaded frequency doubler", Phys. Rev. **A57**, 535 (1998). doi:10.1103/PhysRevA.57.535

[23] M. Jakob, G.Yu. Kryuchkyan, "Photon correlation in an ion-trap system", Phys. Rev. **A59**, 2111 (1999). doi:10.1103/PhysRevA.59.2111

[24] R. Schack, T. A. Brunn, "A C++ library using quantum trajectories to solve quantum master equations", Comp. Phys. Commun. 102, 210-228, (1997). doi:10.1016/S0010-4655(97)00019-2

[25] S. M. Tan, "A computational toolbox for quantum and atomic optics", J. Opt. B1, 424, (1999). doi:10.1088/1464-4266/1/4/312

[26] Vukics, A., Ritsch, H., "C++ QED: an object-oriented framework for wave-function simulations of cavity QED systems", The European Physical Journal D 44, (2007). doi:10.1140/epjd/e2007-00210-x

[27] Zs. Nmeth, G. Dzsa, R. Lovas, P. Kacsuk, "The P-GRADE grid portal", Springer 3044, Lecture Notes in Computer Science, (2004).

[28] H. Astsatryan, Yu. Shoukouryan, V. Sahakyan, "Grid Activities in Armenia", Proceedings of the International Conference Parallel Computing Technologies (PAVT'2009), Novgorod, Russia, (2009).

[29] Operational Portal of European Grid Initiative, http://operations-portal.egi.eu/vo

[30] Hrachya Astsatryan, Vladimir Sahakyan, Yuri Shoukourian, Michel Dayde and Aurelie Hurault, "Enabling Large-Scale Linear Systems of Equations on Hybrid HPC Infrastructures", September 4-16, Skopje, Macedonia, Springer Advances in Intelligent and Soft Computing, 2012, Volume 150/2012, 239-245, DOI: 10.1007/978-3-642-28664-3_22.